\documentclass[twocolumn,reprint,nofootinbib,longbibliography,prd]{revtex4-1}
\usepackage{amsfonts, amsmath, amssymb, bm, enumerate, graphicx, graphics, color,mathrsfs,hyperref}

\newcommand{\Lie}[0]{{\cal L} }

\newcommand{\tl}{\theta_{(\ell)}}

\newcommand{\tN}{\theta_{(N)}}

\newcommand{\nn}{\nonumber}
\newcommand{\be}{\begin{equation}}
\newcommand{\ee}{\end{equation}}
\newcommand{\bea}{\begin{eqnarray}}
\newcommand{\eea}{\end{eqnarray}}

\newcommand{\tq}{\tilde{q}}

\newcommand{\pa}{\partial}

\begin{document}

\title{Collapse and bounce of null fluids}

\date{\today}
\author {Bradley Creelman}
\email{bjc572@mun.ca}
\affiliation{
Department of Mathematics and Statistics\\ Memorial University of Newfoundland \\  
St. John's, Newfoundland and Labrador, A1C 5S7, Canada \\
}
\author{Ivan Booth}
\email{ibooth@mun.ca}
\affiliation{
Department of Mathematics and Statistics\\ Memorial University of Newfoundland \\  
St. John's, Newfoundland and Labrador, A1C 5S7, Canada \\
}

\begin{abstract}
Exact solutions describing the spherical collapse of null fluids can contain regions which violate the energy 
conditions. Physically the violations occur when the infalling matter continues to move inwards even when non-gravitational repulsive forces 
become stronger than gravity. In 1991 Ori proposed a resolution for these violations: spacetime surgery should be used to replace the energy condition violating region
with an outgoing solution. The matter bounces. We revisit and implement this proposal for the more general Husain null fluids including a careful study of 
potential discontinuities and associated matter shells between the regions. 
Along the way we highlight an error in the standard classification of energy condition violations for Type II stress-energy tensors. 
\end{abstract}

\maketitle

%
%
%
%
%
%
%
%

\section{Introduction}

\begin{figure}
\includegraphics{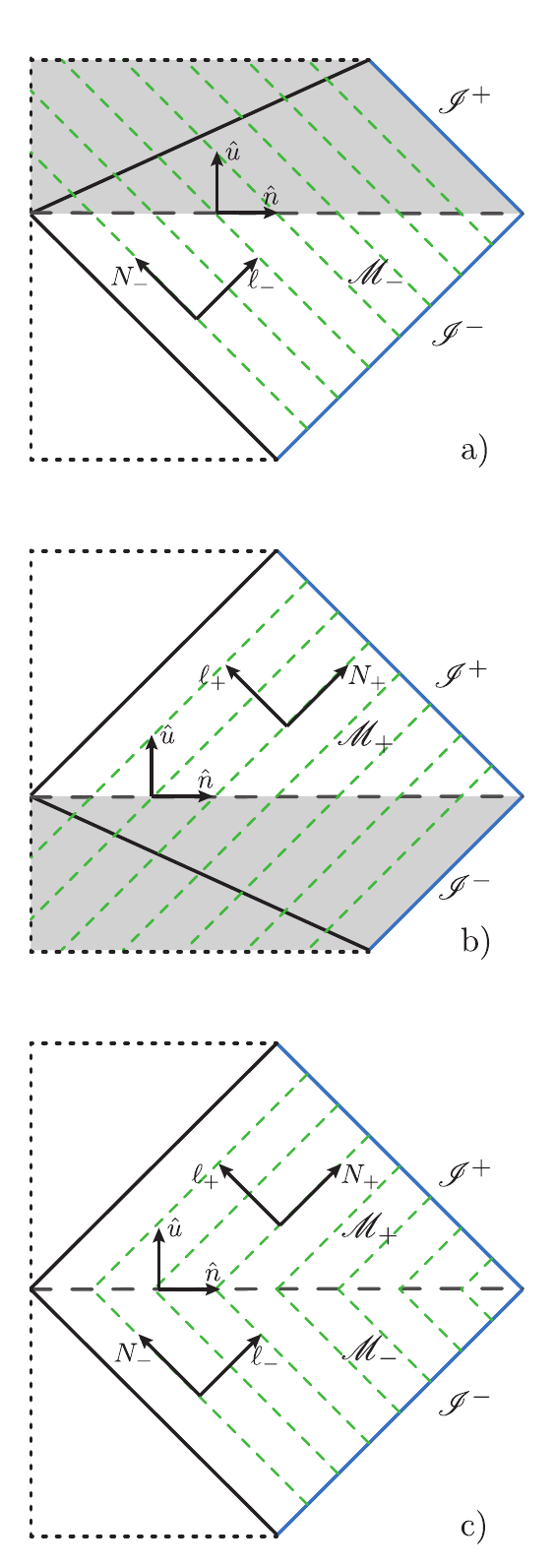}
\caption{Surgery to remove energy condition violating regions from Vaidya Reissner-Nordstr\"om. Subfigure a) shows infalling Vaidya RN (dust accreting onto an RN black hole)  while subfigure b) shows outgoing Vaidya RN (dust being emitted from an RN white hole). In both cases energy condition violations exist in the shaded region. However, as shown in c) if energy condition violating 
regions of each spacetime are removed along the spacelike dashed line, then the remaining pieces may be reconnected into a non-energy condition violating spacetime (apart from maybe at the
junction -- see section \ref{shell}). In all diagrams, apparent rather than event horizons are shown and so in regions where matter crosses the horizon they are spacelike (see, for example, 
\cite{Ashtekar:2004cn,Booth:2005qc}). The dotted lines on all figures indicate that 
they continue in those directions (but into regions that are not of direct interest for us). Note that in the violation-free spacetimes, matter crosses neither the black hole nor white hole horizon.}
\label{mfig}
\end{figure}

Vaidya spacetimes are the best known exact solutions describing  dynamical black (or white) holes. The basic solution describes a null dust infalling onto a black hole (or radiating from a white hole) and 
was later generalized to charged null dust in \cite{Bonnor:1970zz} and to a null fluid with pressure in \cite{husain:1996}. Focussing on collapse solutions, the inclusion of these extra interactions
can result in regions where the energy condition are violated (see, for example,  \cite{ori1991, Lake:1991, kaminaga, husain:1996}). For collapsing matter, these are regions  where the fluid continues moving inward despite non-gravitational repulsive forces becoming stronger than the  gravitational attraction (FIG.~\ref{mfig}a).

For the case of the charged Vaidya solution, Ori\cite{ori1991} argued for a construction to remove the apparent violations.  By 
carefully considering the Lorentz force on the dust and thus including a Lorentz force term in the equations of motion, he showed that on
the hypersurface dividing  regular spacetime from  the region of violations, the wave-vector of the fluid vanishes. 

This suggested a physical reinterpretation of charged Vaidya in which the vanishing wave-vector hypersurface signals a bounce from infalling to outgoing dust. Geometrically
this reinterpretation corresponds to a new hybrid spacetime built from violation-free regions of infalling and outgoing Vaidya solutions (FIG.~\ref{mfig} again). These regions 
join along a common spacelike bounce surface\footnote{This is not a physical restriction but rather based on the available solutions. A timelike bounce would necessarily include 
regions with both infalling and outgoing dust but we do not have an exact solution describing this situation. Hence the construction  can only be used to describe spacelike bounces. }.

This bounce resolves the energy condition violations with the critical hypersurface  corresponding physically to the location where the Lorentz repulsive force overcomes gravity and the 
charged fluid turns around. This interpretation is consistent with the null limit for timelike fluids\cite{ori1991} as well as the evolution of null charged particles in Reissner-Nordstr\"om (RN) spacetimes
\cite{ori1991} (and Appendix \ref{appNullPaths} of this paper) and null charged thin shells\cite{Dray1990}. 

Generalizations of this procedure have recently been applied to modified $f(R)$ theories of gravity \cite{Chatterjee:2015cyv} as well as the extremal case of the charged Vaidya solution 
\cite{Booth:2015kxa}. However in \cite{Booth:2015kxa} a possible inconsistency was noted in Ori's original calculation. In \cite{ori1991} it was found that the extrinsic curvatures of the 
component spacetimes matched along the junction and so, by the standard Israel-Darmois junction conditions \cite{Israel66}, the connection is smooth to first order. In 
\cite{Booth:2015kxa} it was shown that, at least in the extremal case, the extrinsic curvatures do not match and so a thin shell  discontinuity (that is the instantaneous appearance of a 
stress tensor) is required to connect the spacetimes across the bounce surface. Though this was a very special limiting case, it was in tension with the apparently more general result. 

 In this paper we revisit Ori's construction with two goals. First we generalize to Husain null fluid spacetimes  \cite{husain:1996}. In general these are interpreted as  null fluids 
 with pressure, however they include Vaidya Reissner-Nordstr\"om (VRN) as a special case where the energy density and  pressure are re-interpreted as arising from a Maxwell field. Second, we carefully re-examine the spacetime surgery to determine whether or not there is a thin 
 shell discontinuity. When looking at the more general case of Husain null fluids, we also answer the question as to why there are conflicting results in \cite{ori1991} and \cite{Booth:2015kxa}: it turns out that both are   mathematically correct but  differ due to a choice in how to match along the junction hypersurface.

In general, when matching two spacetimes along a spacelike hypersurface, there will not be a unique way in which the matching can take place. We find that in general there are two distinct ways to match the spacetimes along the bounce surface: a time reflection and a second, more complicated, matching (which in the static case is simply equivalent to the transformation from ingoing to outgoing coordinates). For extremal VRN, only the time reflection is possible and in that case it is intuitively clear that the extrinsic curvatures will be the negatives of each other. However this does not
show that there is also  a thin shell in Ori's case: he used the second matching. In that case the shell vanishes not only for VRN but also for the more general Husain null fluids. 
Thus the two results do not contradict each other.
 
%

Along the way we note another, minor result. Almost all stress-energy tensors studied in this paper are of Type II \cite{hawking73}. 
Since we are  concerned with energy condition violations, we re-examined those conditions and were surprised to find an error in their original 
presentation in \cite{hawking73}. While we have subsequently learned that this has been previously noted (see for example \cite{MMS1996,OriPC,Martin-Moruno:2013sfa,RefPC}) the error is not 
universally known and the incorrect conditions have been and continue to be used in the literature (see, for example, \cite{husain:1996,Wang:1998qx,Harko:2000ni,Ghosh2002,Debnath:2007vb,Ghosh:2008zza,Chatterjee:2015cyv}). As such for future reference we explicitly present the correct form of the energy conditions 
in an appendix to this paper. 

%

The paper is organized as follows.  Section \ref{HNF}  reviews Husain null fluids as 
a generalization of the charged Vaidya solution and discusses energy condition violations in these spacetimes. Section \ref{shell} considers the (non-)existence of a thin shell at a spacelike
junction hypersurface for the Husain spacetimes and examines other possible discontinuities in the matter fields. 
Section \ref{examples} demonstrates a concrete example of the matching conditions  and so confirms that the conditions assumed in the previous section are consistent with real examples.  Section \ref{conc} reviews and discusses implications
of the work. Finally, Appendix \ref{appNullPaths} reviews null particle paths in Reissner-Nordstr\"om spacetimes while  Appendix \ref{EC} studies the energy conditions for
Type II stress-energy tensors. 

For notation, early alphabet latin letters $(a,b,c \dots)$ are used as four-dimensional abstract indices, greek letters $(\alpha, \beta, \gamma \dots )$  are used as four-dimensional coordinate indices and
mid-alphabet latin letters with hats  $(\hat{\imath}, \hat{\jmath}, \hat{k} \dots)$ are used as indices for a three-dimensional orthonormal spacelike triad spanning the tangent space of the junction surface.

%
%

\section{Husain Null Fluids}
\label{HNF}

In this section, we review the geometry and physics of the Husain null fluid spacetimes as presented in \cite{husain:1996} and the occurrence of energy condition violations for these solutions.


\subsection{The Spacetime}

The (infalling) Husain solution is obtained by assuming a general spherically symmetric solution with mass function $m(v,r)$:
\be 
ds^2 = -\left(1 - \frac{2m(v,r)}{r} \right) dv^2 + 2dvdr + r^2 d\Omega^2 \, ,
\ee
where $v$ labels infalling null geodesics and $r$ (the areal radius) is an affine parameter along those geodesics: see FIG.~\ref{mfig}a). 

Then from the Einstein equations, the associated stress-energy tensor may be written relative to radially outward- and inward-pointing null vectors
\begin{align}
\ell &= \frac{\partial}{\partial v} +  \frac{1}{2} \left(1 - \frac{2m(v,r)}{r} \right) \frac{\partial}{\partial r}  \; \; \mbox{and} \\
N &= - \frac{\partial}{\partial r} \, ,
\end{align}
as 
\be 
T_{ab} = \mu N_{a} N_{ b} - \rho q^\perp_{ab} + P \tilde{q}_{ab}, \label{NFstress}
\ee 
where $\tilde{q}_{ab}$ and $q^\perp_{ab}$ are respectively the induced metric on the surfaces of constant $(v,r)$ and that on the normal space to those surfaces:
\be
{q}^\perp_{ab} = - \ell_a N_b - N_a \ell_b \label{qperp}
\ee
and 
\be
\tilde{q}_{ab} = g_{ab} - q^\perp_{ab} \, . 
\ee
Further
\be
\mu = \frac{{m_v}}{4 \pi r^2} \,  , \label{mu}
\ee
\be \label{3}
\rho = \frac{m_r}{4 \pi r^2} 
\ee
and
\be \label{4}
P = -\frac{m_{rr}}{8 \pi r}. \ee
where the subscripts are partial derivatives. 

%
%
%


Interpreting these components of the stress-energy tensor, this is an inward falling, self-interacting null fluid. $\mu$ is the flux of energy in the (inward) $N^\alpha$ direction while $\rho$ is the energy 
density associated with the self-interaction and $P$ is a tangential pressure. 

At this stage $m(v,r)$ is arbitrary, however restrictions on its allowed forms are imposed by the energy conditions as outlined in Appendix \ref{EC}. For individual energy conditions, the restrictions
that we find are not equivalent to those given in \cite{hawking73}, however the restrictions imposed if we require that all four energy conditions hold are equivalent. Requiring that the weak, null, 
dominant and strong all hold for (\ref{NFstress})  we must have:
\be
\mu \geq 0 \; , \; \;  \rho \geq 0 \; \mbox{and} \; \; 0 \leq P \leq \rho \, .  \label{EC1}
\ee
That is
\be
{m_v} \geq 0 \; , \; \; m_r \geq 0 \; \mbox{and}  \; -2m_r \leq r m_{rr} \leq 0 \, .  \label{EC2}
\ee

\subsection{Polytropic fluids}

\subsubsection{ $(a\!=\!1)$-fluid infalling onto a black hole}
Even with the energy condition restrictions the range of allowed forms for $m(v,r)$ is still large. However for any particular null fluid one expects an equation of state to relate at least the 
pressure $P$ and energy density $\rho$. Husain focusses on polytropic fluids for which
\be
P = k \rho^a \, ,
\ee
for some constants $k$ and $a$. For our purposes it will be sufficient to consider fluids for which $a=1$. Other, more complicated, equations of state are considered in \cite{husain:1996}.

$P = k \rho$ yields an integrable equation for the mass function $m(v,r)$ and has the solution 
\be \label{p6}
m(v,r) = M(v) -  \frac{g(v)}{2r^{2k-1}} \, , 
\ee
where $M(v)$ and $g(v)$ are arbitrary functions. That is 
\be
ds^2 =   - \left(1 - \frac{2M(v)}{r} + \frac{g(v)}{r^{2k}} \right)dv^2  + 2dvdr + r^2 d\Omega^2 \, .  \label{ds2}
\ee
We restrict our attention to asymptotically flat spacetimes $k>\frac{1}{2}$. Note in particular that choosing $k=1$ and $g(v) = Q(v)^2$ we recover the charged Vaidya solution.

\subsubsection{Energy Conditions} 
\label{sEC1}

Let us now consider restrictions imposed on these solutions by the energy conditions. First for (\ref{ds2})
\be
\rho = \frac{m_r}{4 \pi r^2} =   \frac{(2k-1)}{8 \pi}\frac{g(v)}{r^{2k+2}} \, . 
\ee
Hence with $k>\frac{1}{2}$,
\be
\rho \geq 0 \; \; \Longrightarrow \; \; g(v) \geq 0 \, 
\ee
and so we can rewrite the line element of the Husain spacetime as 
\begin{align}
ds^2 = &   - \left(1 - \frac{2M(v)}{r} +  \left( \frac{\Xi(v)}{r}\right)^{2k} \right)dv^2 \label{ds2_2}\\
&  + 2dvdr + r^2 d\Omega^2 \, .  \nonumber
\end{align}
where we have rewritten $g(v) = \Xi(v)^{2k}$ so that the free function $\Xi(v)$ will have dimensions of length. For $k=1$ we recover VRN with $\Xi(v) = Q(v)$.

Next
\be
0 \leq P \leq \rho \; \; \Longrightarrow \; \;  k \leq 1
\ee
and so now $k$ is bound both above and below: $\frac{1}{2} < k \leq 1$.

Finally  $\mu \geq 0$ requires
\be \label{9}
 m_v= M_v  - k\Xi_v \left(\frac{\Xi}{r}\right)^{2k-1}  \geq 0 . 
\ee
Unlike other violations this one cannot always be removed by restricting our attention to a subclass of solutions. Defining 
\be
R_o = \Xi \left(k  \left| \frac{\Xi_v}{M_v} \right| \right)^{1/(2k-1)} \, , 
\label{12}
\ee
there are four cases:
\begin{enumerate}
\item $\Xi_v > 0$, $M_v \leq 0  \; \Rightarrow$  violations for all $r$,
\item $\Xi_v > 0$, $M_v > 0 \; \Rightarrow$  violations for  $r < R_o$, 
\item $\Xi_v < 0$, $M_v < 0 \; \Rightarrow$ violations for  $r > R_o$
\item $\Xi_v < 0$, $M_v \geq 0 \; \Rightarrow$  no violations.
\end{enumerate}
As a special case note that if $ \Xi_v \Xi^{2k-1}= 0$ then there are no violations as long as $M_v \geq 0$. These are VRN spacetimes but with uncharged dust.  

For general VRN ($k=1$, $\Xi = Q$) there are clear physical interpretations in analogy with the paths of null particles moving in a background RN field. Those paths are considered in 
some detail in Appendix \ref{appNullPaths} and in making the connection note that the radial evolution of any particular shell of constant $v$ is equivalent to that of a corresponding particle of
with an energy at infinity of 
\be
E_\infty = \frac{M_v}{4 \pi }
\ee
and charge
\be
q = \frac{Q_v}{4 \pi r^2} 
\ee
moving in a background RN spacetime with $Q = Q(v)$ and $M=M(v)$. Then the four cases above respectively map onto cases $I_{+-}$, $I_{++}$, $I_{--}$ and $I_{-+}$ from the appendix. 

The interpretation of $M_v$ as proportional to energy at infinity continues for the $k\neq 1$ cases however the individual particle interpretation is then not so clear. 

%
%

\subsubsection{$(a\!=\!1)$-fluid radiating from a white hole}
\label{sEC2}
Thus far we have considered spacetime with matter infalling onto a black hole, however a judicious application of negative signs switches these solutions to ones with matter radiating from 
a white hole. 

In this case the line element is
\be 
ds^2 =  - \left(1 - \frac{2M(u)}{r} + \left( \frac{\Xi(u)}{r}\right)^{2k} \right)du^2 - 2dudr + r^2 d\Omega^2, \label{radiating}
\ee
where $u$ labels the outgoing radial null geodesics and $r$ is still the affine parameter. The stress-energy tensor still takes the form (\ref{NFstress}) though this time for null vectors:
\begin{align}
\ell &= \frac{\partial}{\partial v} -  \frac{1}{2} \left(1 - \frac{2m}{r} \right) \frac{\partial}{\partial r}  \; \; \mbox{and} \\
N  &=  \frac{\partial}{\partial r} \, . 
\end{align}
$N$ continues to point in the direction of the fluid motion and so in this case outwards rather than inwards.

The $\rho \geq 0$ and $0 \leq \rho \leq P$ conditions are unchanged and for $\mu \geq 0$ 
with 
\be
\tilde{R}_o= \Xi \left(k \left| \frac{\Xi_u}{{M_u}} \right| \right)^{1/(2k-1)} \, . 
\ee
there are the same four cases:
\begin{enumerate}
\item $\Xi_u > 0$, $M_u \leq 0  \; \Rightarrow$  violations for all $r$,
\item $\Xi_u > 0$, $M_u > 0 \; \Rightarrow$  violations for $r < \tilde{R}_o$, 
\item $\Xi_u < 0$, $M_u < 0 \; \Rightarrow$ violations for  $r >  \tilde{R}_o$
\item $\Xi_u < 0$, $M_u \geq 0 \; \Rightarrow$  no violations.
\end{enumerate}
Again the  $k=1$ the cases may be understood in terms of the evolution of charged null particles in an RN background. This time they are the outgoing particles $O_{+-}$, $O_{++}$, $O_{--}$
and $O_{-+}$ discussed in Appendix \ref{appNullPaths}.

\section{Surgery to remove energy condition violations}
\label{shell}

The complementary energy condition violations for infalling and radiating null fluids suggest replicating Ori's construction for these 
more general spacetimes. That is for ${M}_v > 0$ excise the $r< R_o(v)$ section of an infalling spacetime (\ref{ds2})
and replace it with the $r< \tilde{R}_o(u)$ section of a radiating spacetime (\ref{radiating}) with the parameters chosen so that 
the induced metrics match on $r = R_o(v) = \tilde{R}_o(u(v))$ for some function $u(v)$. 

As we shall now see, the Israel-Darmois junction conditions require $\frac{du}{dv} < 0$ along the matching surface. Thus, referencing the lists in Sections \ref{sEC1} and \ref{sEC2},
 these are case $ 2 \leftrightarrow 3$ matchings. 

%
%

\subsection{Hypersurface geometry}
First we study the intrinsic and extrinsic geometry of spherically symmetric hypersurfaces. 


%
%

It will be convenient to consider both infalling and radiating spacetimes simultaneously and so we write 
\be
ds^2 = -f(w,r)dw^2 + 2\epsilon dw dr + r^2 d\Omega^2 \, , \label{genMet}
\ee
where $\epsilon = \pm 1$ with $w=v$ (ingoing) for $\epsilon = 1$ and $w=u$ (outgoing) for $\epsilon = -1$. We leave the metric function
in the general form
\be
f(w,r) = 1 - \frac{2m(w,r)}{r} \, . 
\ee
For this discussion the more specialized form (\ref{ds2_2}) is not required and in fact it is simpler to write our expressions in terms of $f(w,r)$ or $m(w,r)$. 


Now consider a general spherically symmetric hypersurface $B$ parameterized by  $w = W(\lambda)$ and $r = R(\lambda)$. Then the induced metric on $B$ is
\be \label{induced}
d\Sigma^2 =  (-f \dot{W}^2 + 2\epsilon \dot{W}\dot{R})d\lambda^2 + R^2 d\Omega^2 \, , 
\ee
with dots indicating derivatives with respect to $\lambda$. We restrict our attention to spacelike $B$ and so the functions must satisfy
\be
\dot{W} \left( -f \dot{W} + 2\epsilon \dot{R} \right) > 0  \, . 
\ee

Turning to the extrinsic geometry it is convenient to work with a hypersurface-adapted tetrad. The timelike unit normal pointing in the positive $w$ direction is
\be
\hat{u}^{\alpha} \partial_\alpha  \equiv \hat{e}_{(0)}^\alpha \partial_\alpha \equiv \frac{1}{\sqrt{2\epsilon R_{w} - f}}\left( \frac{\partial}{\partial w} +  \left(\epsilon f - R_w \right) \frac{\partial}{\partial r} \right) \, , 
\ee
 and the spacelike unit tangent pointing in the positive $r$ direction is
\be
\hat{n}^{\alpha}\partial_{\alpha} \equiv \hat{e}_{(1)}^\alpha \partial_\alpha \equiv \frac{\epsilon}{\sqrt{2\epsilon R_{w} - f}}\left(\frac{\partial}{\partial w} + R_{w}\frac{\partial}{\partial r}\right) \, . 
\ee
In both cases  $R_w \equiv \frac{\dot{R}}{\dot{W}} = \frac{dR}{dw}$ if we reparameterize $B$ as  $r=R(w)$.  
Finally the tangential unit vectors are
\be
\begin{split}
&{\hat{e}_{\theta}}^{\alpha} \pa _{\alpha} \equiv {\hat{e}_{(2)}}^{\alpha} \pa _{\alpha} \equiv \frac{1}{r}\frac{\partial}{\partial \theta} \; \; \mbox{and} \\
&{\hat{e}_{\phi}}^{\alpha}\pa_{\alpha} \equiv {\hat{e}_{(3)}}^{\alpha}\pa_{\alpha} \equiv \frac{1}{r \sin\theta} \frac{\partial}{\partial \phi} \, . \\
\end{split}
\ee

The  extrinsic curvature of $B$ relative to the tetrad is
\be
K_{\hat{\imath}\hat{\jmath}} = \hat{e}_{\hat{\imath}}^\alpha  \hat{e}_{\hat{\jmath}}^\beta \nabla_\alpha \hat{u}_\beta \, . 
\ee
That is
\begin{align}
K = &\left( \frac{ - \epsilon (2 R_{ww} + ff_r) + ( f_w  +  3 f_r R_{w})  }{2\left(2\epsilon R_{w} - f\right)^{3/2}} \right)  \left(\hat{n} \otimes \hat{n} \right)  \nonumber \\
 & +   \left( \frac{ \epsilon f -  R_w }{r  \sqrt{2\epsilon R_{w} - f}}  \right) \left(\hat{e}_{\theta} \otimes \hat{e}_{\theta} + \hat{e}_{\phi} \otimes \hat{e}_{\phi} \right) \, . \label{K}
\end{align}
Subscripts indicate (partial) derivatives: $f_r = \partial_r f$, $f_w = \partial_w f$ and $R_{ww} = \frac{d^2}{dw^2} R(w)$. 


Finally note that relative to the hypersurface tetrad
\begin{align}
\ell &= \frac{\sqrt{2 \epsilon R_w - f}}{2} \left(\hat{u} + \epsilon \hat{n} \right)  \; \; \mbox{and} \label{lho}\\
N &= \frac{1}{\sqrt{2 \epsilon R_w - f}} \left(\hat{u} - \epsilon \hat{n} \right) \label{Nhor}\, . 
\end{align}
The spacelike tangent vector $n$ always points in the positive-$r$ direction but $\ell$ and $N$ are instead tied to the fluid flow and so change orientations depending on whether we are considering the infalling or radiating
solution.

\subsection{Matching infalling and radiation solutions across $B$: geometry}

Now consider $B$ embedded into both an infalling spacetime $\mathcal{M}_-$ and a radiating spacetime $\mathcal{M}_+$ (the subscript indicates that in the final 
construction $\mathcal{M}_-$ will be in the past of $\mathcal{M}_+$ as in FIG.~\ref{mfig}).  Parameterize the two embeddings as:
\be
(v,r) = (V(\lambda), R(\lambda)) \; \; \mbox{and} \; \; (u,r) = ({U}(\lambda), \tilde{R}(\lambda)).
\ee
We then restrict our attention to matchings for which
\be
f^-(U(\lambda), R(\lambda)) = f^+(V(\lambda), R(\lambda)) \, . \label{fA}
\ee
While it may be possible to construct matchings for more general surfaces, this is both computationally convenient and gives rise to solutions with desirable physical properties (Section \ref{mattermatching}). 
%

\subsubsection{Matching the induced metric}
Matching the components of the induced metrics (\ref{induced}) on $B$, the angular components give
\be
R(\lambda) = \tilde{R}(\lambda) 
\ee
and so henceforth we discard the tilde. The $(\lambda, \lambda)$ components give  
\be
f  \dot{V}^2 - 2 \dot{V} \dot{R} = f \dot{U}^2 + 2 \dot{U} \dot{R} \label{llcomp} \, 
\ee
where we have omitted the superscripts to distinguish the $f$s since they agree on $B$. 
Then the induced metrics match if
\be
 \left( \dot{V} + \dot{U} \right) \left(f (\dot{V} - \dot{U} ) - 2 \dot{R} \right) = 0 \, .  
\ee
Thus there are two possible matchings\footnote{Equivalent matchings have been previously been discussed in \cite{Prats1995,Fayos:1996gw,Fayos:2002rm} for matching spherically symmetric spacetimes along a surface of arbitrary signature. } which we label as
\begin{align}
 \mbox{\bf Reflective:  }  \dot{U} = - \dot{V} &  \; \; \Longrightarrow \; \; U_v = - 1  \label{refsym} \; \; \mbox{and} \\
 \mbox{\bf Ori:  }  \dot{U} = \dot{V} - \frac{2\dot{R}}{f} &  \; \; \Longrightarrow \; \; U_v = 1 - \frac{2R_v}{f}   \, . 
\label{orichoice}
\end{align}
where the right-hand expressions arise if we adopt the ingoing $v$ as our surface parameter: $\lambda=v$. Henceforth we make this choice. Note that 
in both cases $U_v < 0$. 

As suggested by the label, the first solution (\ref{refsym}) corresponds to a time-reversal symmetry between the regions: $K^+_{\hat{\imath}\hat{\jmath}}= - K^{-}_{\hat{\imath}\hat{\jmath}}$.
This is the matching condition that was used in \cite{Booth:2015kxa}. However the second solution (\ref{orichoice}) is the one that was used by Ori. For pure Schwarzschild or RN this is just the 
transformation that re-parameterizes the surface from ingoing to outgoing coordinates. 

Given that different matchings were being used the disagreement between the papers is not surprising! Note however that this was 
unavoidable as in  \cite{Booth:2015kxa} the matching was along the apparent horizon where $f=0$ and so Ori's choice was not available (or noticed by the author). 

\subsubsection{Matter shell from matching the  extrinsic curvatures}

For either of these choices, we can apply the Israel-Darmois junction conditions \cite{Israel66} to calculate the stress-tensor necessary to account for any  discontinuities introduced by the construction. 
Recall that if the extrinsic curvatures of $B$ in $\mathcal{M}_-$ and $\mathcal{M}_+$ are not equal then $K^-_{\hat{\imath}\hat{\jmath}} \neq K^+_{\hat{\imath}\hat{\jmath}}$ and there is a thin shell of matter at $B$ with stress-tensor
\be \label{2}
S_{\hat{\imath} \hat{\jmath}} = -\frac{1}{8\pi}\left([K_{\hat{\imath}\hat{\jmath}}]  -  [K] h_{\hat{\imath}\hat{\jmath}} \right),
\ee
where 
\be
[K_{\hat{\imath}\hat{\jmath}}]  = K^+_{\hat{\imath}\hat{\jmath}} - K^-_{\hat{\imath}\hat{\jmath}}
\ee
and similarly $[K] = h^{\hat{\imath}\hat{\jmath}} [K_{\hat{\imath}\hat{\jmath}}]$. Then the radial and tangential pressure densities are respectively
\begin{align}
S_{\hat{n} \hat{n}} &  = \frac{1}{4 \pi} [K_{\hat{\theta} \hat{\theta}}] \; \; \mbox{and} \; \;  \\
S_{\hat{\theta} \hat{\theta} } = S_{\hat{\phi} \hat{\phi} } & = \frac{1}{8 \pi} \left( [K_{\hat{\theta} \hat{\theta}}] +[ K_{\hat{n} \hat{n}}   ] \right)  \, . 
\end{align}
These components can be calculated from (\ref{K}):
\be
[K_{\hat{\theta} \hat{\theta}}] = \frac{2 R_v - f (1 - U_v ) }{ R \sqrt{2R_v - f}}
\ee
is easy while
\begin{align}
[K_{\hat{n} \hat{n}}] =&   \frac{2 R_{vv} \left(1 - U_v \right) + 2 R_v U_{vv} + f f_r (1-U_v^3)}{2 (2 R_v - f)^{3/2}}  \\
& - \frac{ (f_v + 3 f_r R_v) (1+U_v^2)}{2 (2 R_v - f)^{3/2}}   \nonumber
\end{align}
is more complicated. In both of these calculations we have eliminated $R_u$ using  $R_u = \frac{R_v}{U_v}$ in the numerators and 
\be
 \frac{1}{\sqrt{-f - 2R_{u}}}  = -  \frac{U_v}{\sqrt{-f + 2R_{v}}}\, , 
\ee
which can be derived directly from (\ref{llcomp}), for denominators.

Now we specialize to the reflective and Ori matchings. For reflective $U_{vv} = 0$ and so
\begin{align}
S^{\mbox{\tiny{ref}}}_{\hat{n} \hat{n}} & = \frac{R_v - f}{2 \pi R \sqrt{2R_v - f}} \; \; \mbox{and} \label{Snnref} \\
S^{\mbox{\tiny{ref}}}_{\hat{\theta} \hat{\theta}} & = \frac{R_v - f}{4 \pi R \sqrt{2R_v - f}} + \frac{2 R_{vv}  + f_r (f - 3 f_r R_v) - f_v}{8 \pi (2 R_v-f)^{3/2}} \label{S22ref} \, .
\end{align}
while for Ori 
\be
U_{vv} = - \frac{2R_{vv}}{f} + \frac{2 f_rR_v^2}{f^2} + \frac{2 f_v R_v}{f^2}
\ee
and so 
\begin{align}
S^{\mbox{\tiny{Ori}}}_{\hat{n} \hat{n}} & =0 \; \; \mbox{and} \label{SnnOri} \\
S^{\mbox{\tiny{Ori}}}_{\hat{\theta} \hat{\theta}} & =     \frac{f_v}{4 \pi R \sqrt{2R_v - f}} \label{S22Ori} \, . 
\end{align}

Thus far $B$ has been  a general spacelike surface. However we are mainly interested in surfaces for which the matching is as smooth as possible and so now 
restrict our attention to  $B$ defined by 
\be
\mu = 0 \Leftrightarrow f_v = 0 \, . 
\ee
Then by the discussion surrounding  (\ref{mu}) the flow of energy in the ingoing null direction vanishes at $B$ and can continuously switch from ingoing to outgoing.

That this physically motivated choice is achievable is demonstrated in Section \ref{examples}, but for now we note that with $f_v = 0$,
\be
S^{\mbox{\tiny{Ori}}}_{\hat{\imath} \hat{\jmath}} = 0
\ee
while the reflective matching retains non-zero components. 

In a little more detail for the polytropic fluid and a reflective matching:
\be
f^{\mbox{\tiny{p,r}}}_v = 0 \; \; \Longleftrightarrow \; \;  \left( \frac{\Xi}{r} \right)^{2k} = \frac{\Xi}{kr} \frac{1}{\frac{M_v}{\Xi_v}}
\ee
and we can apply the right-hand side equality to show that
\be
f^{\mbox{\tiny{p,r}}}_r = \frac{2M}{R^2} \left(1 - \frac{\Xi}{M} \frac{M_v}{\Xi_v}  \right) \, . \label{f__r}
\ee
In particular note that if $\Xi(v) = \xi M(v)$ for some constant $\xi$, then $f_r$ vanishes as well (but is still not sufficient to cause the reflective stress-tensor to vanish). 
 We will return to this in section \ref{linmat}.

\subsection{Matching infalling and radiation solutions across $B$: matter fields}
\label{mattermatching}

We can also consider potential discontinuities in the matter fields across $B$, apart from the shell. We begin by considering jumps in the bulk stress-energy tensor. 

\subsubsection{Discontinuities in the bulk $T^{\mbox{\tiny{bulk}}}_{ab}$}
\label{DTab}
From the standard junction condition formalism, the stress-energy tensor for the full spacetime is
\be
T_{ab} = \Theta^- T^-_{ab} + \delta_B S_{ab} + \Theta^+ T^+_{ab}
\ee
where $\Theta^{\pm} = 1$ on $\mathcal{M}_\pm$ but vanishes on $\mathcal{M}_\mp$ and $\delta_B$ is a Dirac delta function centred on $B$. Thus even if there is no thin shell induced 
on $B$ it is still possible to have a discontinuity in the bulk stress-energy across $B$. The canonical example of such a jump is across the boundary separating the FRW from Schwarzschild region 
during Oppenheimer-Snyder collapse \cite{Oppen}. 

To see if there is such a discontinuity in our case we compare the limiting behaviour of $T^{\mbox{\tiny{bulk}}}_{ab}$ as we approach $B$ from 
the $\mathcal{M}_-$ and $\mathcal{M}_+$ sides. These limits are easily calculated as the fields are continuous up to $B$. For $\mu=0$ one can apply 
(\ref{NFstress}), (\ref{lho}) and (\ref{Nhor}) to find that at $B$:
\begin{align}
T^{\mbox{\tiny{bulk}}}_{ab} = & \rho \hat{u}_a \hat{u}_b - \rho  \hat{n}_a \hat{n}_b + P \tilde{q}_{ab} \, . 
\end{align}

It is straightforward to see that $\rho^+ = \rho^-$ on $B$. From $f^+ = f^-$:
\begin{align}
\frac{d}{d\lambda} \left( f^+ (U(\lambda), R(\lambda)) \right)  = \frac{d}{d\lambda} \left( f^- (V(\lambda), R(\lambda)) \right) \, ,
\end{align}
but since $f^-_u = f^+_v = 0$ (from $\mu = 0$) and $\dot{R} \neq 0$ this implies that 
\be
f^+_r  = f^-_r  \Leftrightarrow  \rho^+ = \rho^- \, ,
\ee
and the only possible stress-energy discontinuity is from the pressure:
\be
\Delta T_{ab} = (P^+-P^-) \tilde{q}_{ab} \, . 
\ee
For the special case of a polytropic null fluid $P=k \rho$ and so there is no discontinuity in the stress-energy tensor. 


The easiest way to do a $f^+=f^-$ match for such a fluid is to require
%
\be
M^- (V(\lambda)) = M^+ (U(\lambda)) \; \; \mbox{and} \; \; \Xi^- (V(\lambda)) = \Xi^+ (U(\lambda)) \,  \label{SpecMod} \, . 
\ee
This is also a physically convenient choice: for this matching when a shell bounces it will return to infinity with the same energy density $M$ (and $\Xi$) as when it left.

\subsubsection{Other discontinuities}
It is possible to have discontinuities in fields that do not show up in either the boundary or bulk stress-energy. For example, discontinuities in the electric field can signal the existence of thin shells
of charge. This is a standard result from undergraduate electromagnetism but as an example in general relativity\footnote{For further discussion of these matching conditions in spherically symmetric general relativity see, for example, \cite{Fayos:1996gw,Fayos:2002rm}.} consider two Reissner-Nordstr\"om spacetimes of the same mass 
but opposite charged attached across an $r=\mbox{constant}$ surface. The geometry is indifferent to the sign of the charge. In particular both the metric and stress-energy depend 
only on the square of the charge:
\begin{align}
f & = 1 - \frac{2M}{r} + \frac{Q^2}{r^2} \\
T_{ab} & = \frac{Q^2}{8 \pi r^4} (-q^\perp_{ab}+ \tilde{q}_{ab})  \, . \label{TEM}
\end{align}
However in this case the stress-energy tensor is generated by the underlying Maxwell field:
\be
F_{ab} = \frac{Q}{r^2} (\ell_a N_b - N_b \ell_a)  \label{FEM}
\ee
where $\ell$ and $N$ are cross-normalized radial null vectors $\ell \cdot N = -1$ and $E_\perp = \frac{Q}{r^2}$ is the radial component of the electric field that can be integrated over
surfaces of constant $r$ to obtain the contained charge $Q$. Hence if $Q_{\mbox{\tiny{in}}} = - Q_{\mbox{\tiny{out}}}$, then even though there are no geometric or stress-energy
discontinuities there is an induced (total) charge of $2 Q_{\mbox{\tiny{in}}}$ at the interface.  

One could ask if similar discontinuities can arise in our more general models. The answer is no -- unless there is an underlying theory generating the $\mu$, $\rho$ and $P$. 
Without an additional theory, all that there is is the stress-energy tensor and so if that is continuous\footnote{In fact it is not hard to see that polytropic fluids the components are actually $C^1$ and so overachieve this target. See Appendix \ref{C1}.}, that is the end of the story. This is the situation for our general models 
except for the polytropic fluid with $k=1$. There one can either: 1) take  $\mu$, $\rho$ and $P$ at face value as the energy densities and pressure associated with a null fluid 
or 2) reinterpret $\mu$, $\rho$ and $P$ as arising from charged null dust. From the perspective of the stress-energy tensor this distinction is irrelevant however taking the Maxwell
interpretation opens the possibility of an electric field discontinuity as discussed above. 

Even with the null dust-Maxwell interpretation and so VRN spacetimes there is no discontinuity for the (\ref{SpecMod}) cases:  if $Q^+ = Q^-$ and the metrics match then so 
do the normal components of the electric field. There is no thin shell of charge.

\section{Bouncing Null Fluid Example}
\label{examples}

We have now seen several properties of the matching surface but have not yet established whether there is any $m(v,r)$ for which it exists with the properties that we have assumed. 
For example is it actually possible to pick $m(v,r)$ so that $\mu = m_v = 0$ is spacelike 
and the surface is not inside a trapped region? In this section we demonstrate at least one example exists. 

\subsection{Trapped and untrapped regions}

First we establish the location of the trapped regions in our spacetimes. For (\ref{genMet}) the outward and inward null expansions are
\be
\tl = \tq^{ab} \nabla_a \ell_b = \frac{\epsilon f}{r} 
\ee
and 
\be
\tN = \tq^{ab} \nabla_a N_b = -\frac{2 \epsilon }{r} \, .
\ee

FIG.~\ref{mfig}a) shows spacetime with $\epsilon = 1$ and an infalling fluid. In that case spherical surfaces are outer trapped  ($\tl<0$) for $f<0$, marginally outer trapped ($\tl =0$) for 
$f=0$ (that is an apparent horizon), and outer untrapped ($\tl > 0$) for $f>0$. For all of these $\tN < 0$ and so when $f<0$ the surfaces are fully trapped and so inside a black hole. 

By contrast  FIG.~\ref{mfig}b) with $\epsilon=-1$ shows a radiating white hole spacetime. The apparent horizon is again at $f=0$ but this time the shaded region is totally untrapped (
$\tl>0$, $\tN > 0$) when $f<0$. So again the region of regular spacetime has $f>0$.

\subsection{Linear matter}
\label{linmat}

Thus the surface $\mu = 0$ is spacelike and always outside of the black and white hole regions if there is a choice of 
 $M(v)$ and $\Xi(v)$ such that both 
\be
2 \epsilon R_w - f(w,R(w))  > 0\; \; \mbox{and} \; \;  
 f(w,R(w)) > 0 \, , 
 \ee
where $R(w)$ is implicitly defined by $f_w(w,R(w)) = 0$. 

To see that these conditions can be met, consider the simple choice
\be
\Xi(w) = \xi M(w)
\ee
where $\xi > 0 $ is a constant. In the charged Vaidya case, $\xi$ corresponds to the charge-to-mass ratio of the fluid. 
With this choice (\ref{12}) gives that the junction hypersurface is
\be  \label{Rlin}
R(w)  = \chi M(w), 
\ee 
where $\chi = \xi \left(k \xi\right)^{1/(2k-1)}$. 

\begin{figure} 
\includegraphics[scale=0.5]{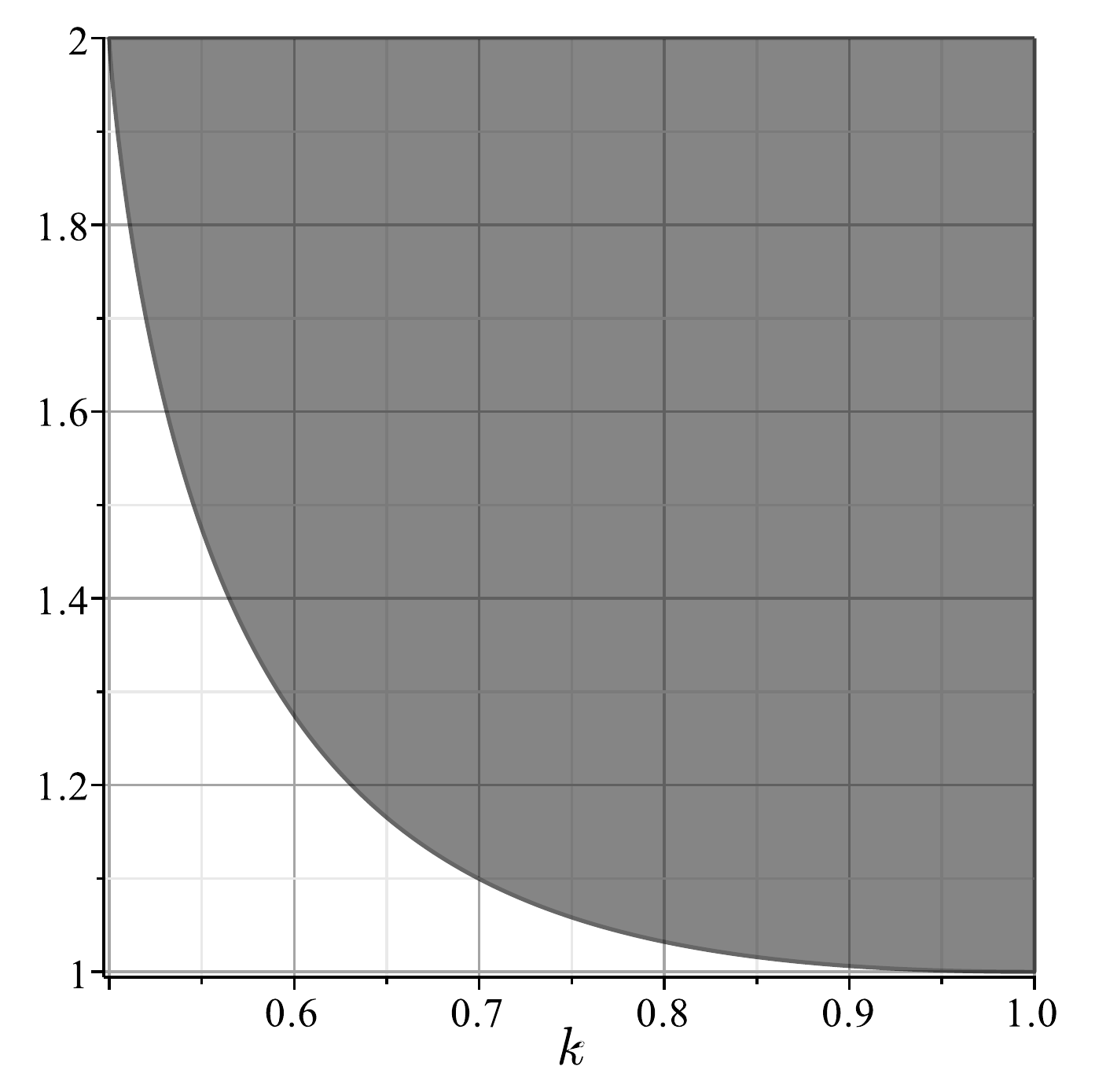}
\caption{Allowed values of $\xi$  so that the surface $m_w=0$ will be between the horizon and $r=\infty$. $\xi$ must be chosen in the shaded region above 
 $\frac{1}{k}(2k-1)^{\frac{2k-1}{2k}}$.  }
\label{xibnd}
\end{figure}

Then
%
%
\be \label{39}
\begin{split}
f(w,R(w)) &= 1 - \frac{2}{\chi } + \left(\frac{\xi}{\chi } \right)^{2k}  \\
& = 1 - \frac{1}{\chi} \left(\frac{2k-1}{k} \right) \\
\end{split}
\ee
is constant along the surface and 
is positive (the shaded region of FIG.~\ref{xibnd}) if
\be
\chi  >  \frac{2k-1}{k} \; \Longleftrightarrow \xi > \frac{(2k-1)^{\frac{2k-1}{2k}}}{k} \, . \label{xib}
\ee
In particular note that for the $k=1$ charged Vaidya case, this condition implies that the charge to mass ratio must be greater than 1. 
The special $(k=1,\xi = 1)$ case is the dynamical extremal horizon considered in \cite{Booth:2015kxa}.

\begin{figure} 
\includegraphics[scale=0.25]{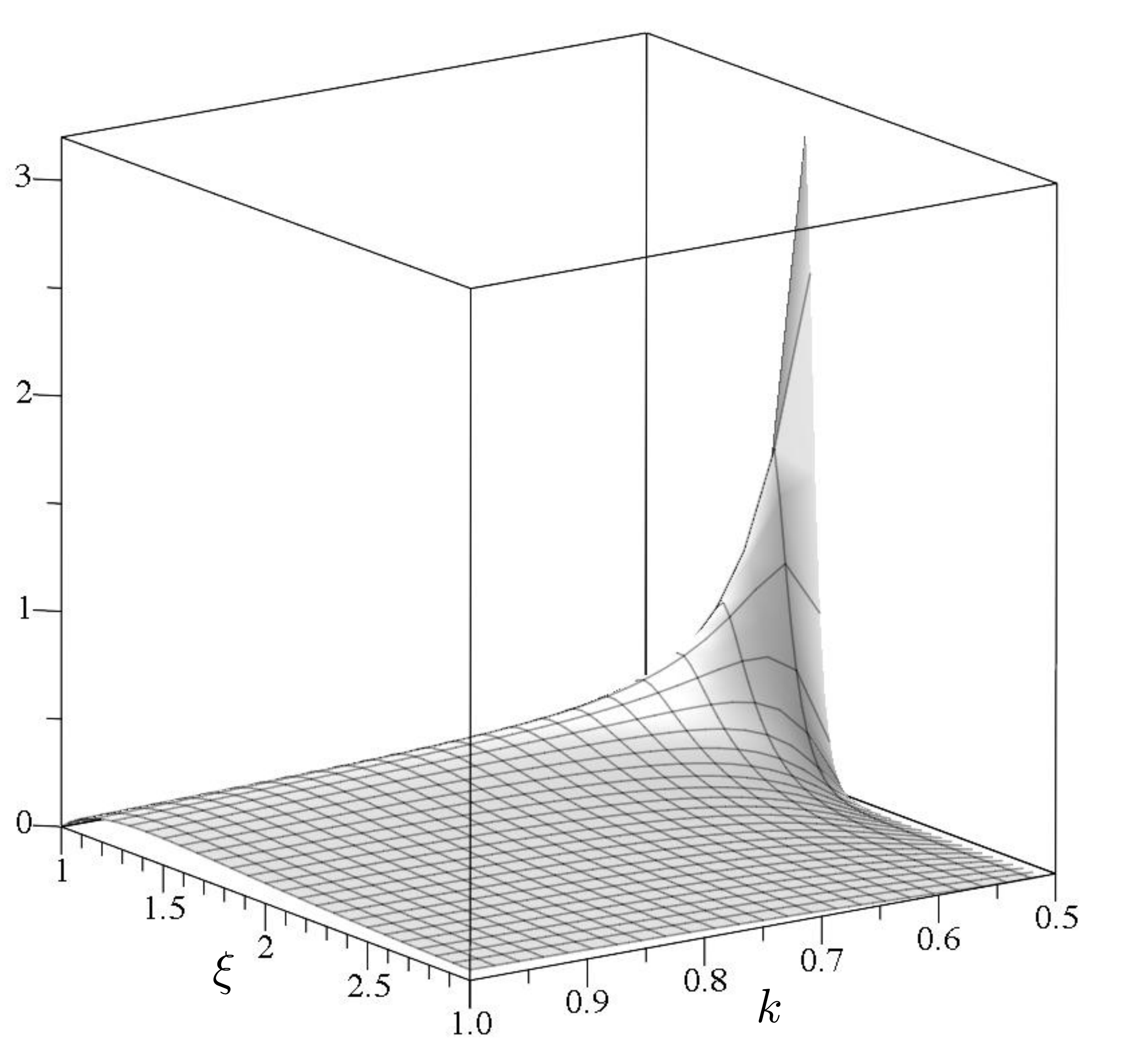}
\caption{Minimum values of $M_v$ for infalling null fluids. $M_v$ must be greater than the plotted surface for the junction surface to be spacelike. For the radiating side $-M_v$ must 
be greater than this value.  }
\label{dMbnd}
\end{figure}

Next, the junction surface  in the ingoing spacetime is spacelike if $2R_v \geq F$. That is on applying (\ref{Rlin}):
\be
M_v > \frac{f}{2\chi} = \frac{1}{2k \chi^2} \left( k \chi + 1 - 2k \right)  \, . 
\ee
Thus for any choice of $(k,\xi)$ there is a lower bound on $M_v$. Equivalently this is a lower bound on the allowed fluid energy at infinity (Appendix \ref{appNullPaths}).
Because we have restricted our attention to junction surfaces outside the black hole this lower
bound is necessarily positive: that is there is a minimum allowed rate of expansion. Similarly in the radiating region there is a minimum allowed rate of contraction. 
This minimum is shown in FIG.~\ref{dMbnd}.

In the extreme Vaidya limit ($k \rightarrow 1, \xi \rightarrow 1$) this bound goes to zero but in all other cases it is positive. As such these constructed spacetimes can only describe 
continuous (eternal) expansions. They cannot describe spacetimes which either depart from or return to equilibrium. However once again it is important to emphasize that this isn't a 
restriction on the allowed physics of spacetimes but rather a restriction on which spacetimes can be described by this particular  model.

%
%
%
%
%

We now examine the stress-energy at a reflective junction for this linear matter. Since both $f_v$ and, by (\ref{f__r}), $f_r$ vanish  the expressions become quite simple: 
\be
S^{\mbox{\tiny{ref}}}_{\hat{n} \hat{n}}  = \frac{ M_v -f/\chi}{2 \pi \chi^{1/2} M \sqrt{2 M_v - f/\chi}}
\ee
where $f$ is given by (\ref{39}). Meanwhile from from (\ref{S22ref}) we have
\be
S^{\mbox{\tiny{ref}}}_{\hat{2} \hat{2}}  = \frac{1}{4 \pi \chi^{1/2}} \left(  \frac{M_v - f/\chi}{M  \sqrt{2M_v - f/\chi}} + \frac{ M_{vv} }{ (2 M_v-f/\chi )^{3/2}} \right)
\ee

For the special case of linear accretion $M = ({f}/{\chi}) v$ both of these vanish, but in general that is not the case.

\section{Conclusion}
\label{conc}

In this paper we have extended Ori's resolution of VRN energy condition violations to Husain null fluids. We saw that  with his matching condition, the no-thin-shell bounce result extends to the 
Husain null fluids. The bounce is naturally caused by the fluid pressure. 

By contrast for the reflective matching conditions used in \cite{Booth:2015kxa}, apart from a very special choice of the parameter functions, there continues to be a thin shell 
at the bounce surface. This is the physical cause of that bounce: it provides the necessary energy to turn the matter around. 
However note that it in itself can  be interpreted as violating the energy conditions: it is 
pressure without a corresponding energy density. 

We also examined  the bulk stress-energy tensor and identified the conditions under which there are discontinuities in the bulk stress-energy tensor at $B$. For 
polytropic fluids with the most convenient matching conditions, the stress-energy tensor and its first derivatives are continuous across the transition. For the special case of 
VRN where the stress-energy tensor is interpreted as arising from a null dust-Maxwell system, there is no thin shell of charge on $B$. 

Finally, we explicitly demonstrated the existence of parameter choices ($\Xi(v) = \xi M(v)$) such that the bounce surface is spacelike and outside of any trapped region. However we also saw that these choices 
restrict us to describing cases where $M_v(v)$ is always greater than some positive constant. Thus 
it necessarily describes an eternally expanding junction surface. While this particular ansatz of solutions cannot describe departures from or returns to equilibrium it still serves to establish
the existence of solutions for which our matching conditions apply.

\section*{Acknowledgements} 
Thanks to John Bowden, Hari Kunduri, Viqar Husain, Matt Visser (in particular for his input on Type II energy conditions) and Jessica Santiago for useful discussions at various points during this work. 
Amos Ori helped us find a significant error in the first version of this paper.
Substantial parts of the calculations were done while IB was
on sabbatical at Victoria University in Wellington, NZ and he would like to thank the School of Mathematics and Statistics for their hospitality. 
IB and BC were both supported by NSERC Discovery Grant 
261429-2013 while BC was also supported by an NSERC PGS-M Fellowship. 

\appendix

\section{Charged null particle paths in Reissner-Nordstr\"om}
\label{appNullPaths}

A charged timelike particle moving in a spacetime with electromagnetic field does not move along geodesics but instead with unit four-velocity $\hat{v}^a$ which obeys
\be
\hat{v}^a \nabla_a \hat{v}^b =  \frac{q}{m} F^b_{\phantom{b}c} \hat{v}^c 
\ee
where $q$ and $m$ are respectively its charge and mass. Similarly Ori\cite{ori1991} argued that the (null) ``wave-vector'' $k^a$ of a massless particle should obey
\be
{k}^a \nabla_a k^b =  q F^b_{\phantom{b}c} k^c  \label{chargedpath}
\ee
where $q$ is again the charge. The scaling of the null vector is significant as an observer with unit four-velocity $u^a$ would measure it to have energy $E= - k \cdot u$. In particular 
we will find it useful to label these paths by their energy observed by an observer at infinity $E_\infty = - k \cdot u_\infty$. 

We study the evolution of charged null particles in RN spacetime. We restrict our attention to particles moving radially and so while we already know that they must follow the 
same paths as null geodesics (\ref{chargedpath}) will fix the scaling of the null vectors. We work with RN in ingoing Eddington-Finkelstein coordinates:
\be
ds^2 = - f dv^2 + 2 dv dr + r^2 d \Omega^2 \, , 
\ee
where $f = 1 - \frac{2M}{r} + \frac{Q^2}{r^2}$ in the usual way but unlike in the main text $M$ and $Q$ are constants. The associated electromagnetic field is generated by the potential
\be
A = - \frac{Q}{r} dv  \; \; \Rightarrow \; \; F = - \frac{Q}{r^2} dv \wedge dr \, .  
\ee
We work with a null dyad of the same form as in the main text:
\begin{align}
\ell & = \frac{\partial}{\partial v} + \frac{f}{2} \frac{\partial}{\partial r} \;  \; \mbox{and} \\
N & = - \frac{\partial}{\partial r} \, . 
\end{align}

We consider ingoing and outgoing particles whose wave-vectors necessarily take the form: 
\be 
k^- = g^-(r) N \; \; \mbox{and} \; \;  k^+ = g^+(r) \ell
\ee
for some functions $ g^-(r) $ and  $g^+(r) $ respectively. By (\ref{chargedpath})
\begin{align}
g^- (r) & = E^-_\infty  - \frac{qQ}{r} \; \; \mbox{and} \\
g^+ (r) & =  \frac{2}{f} \left(E^+_\infty -  \frac{qQ}{r}  \right)  \, . 
\end{align}
Thus the observer hovering at constant $r$ with four-velocity
\be
u = \frac{1}{\sqrt{f}} \frac{\partial}{\partial v} 
\ee
measures energies
\begin{align}
E^- & =  \frac{1}{\sqrt{f}} \left( E^-_\infty  - \frac{qQ}{r}  \right) \\
E^+ & = \frac{1}{\sqrt{f}} \left( E^+_\infty  - \frac{qQ}{r}  \right)
\end{align}
with $E_\infty$ clearly being the limit as $r \rightarrow \infty$. 

\begin{figure*}
\includegraphics{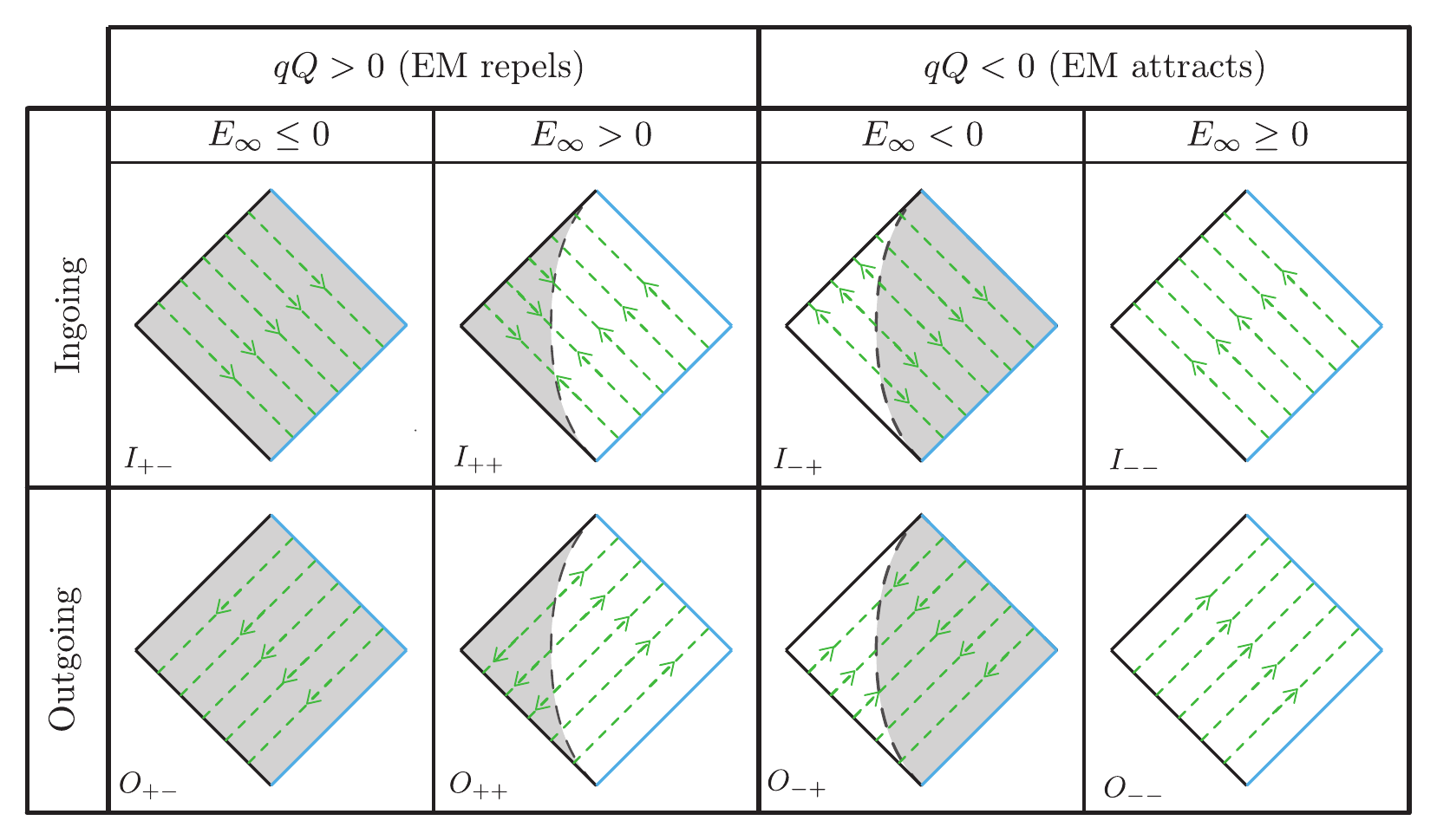}
\caption{Possible charged null particle paths in RN spacetimes. Only the region between the black and white hole event horizons and null is shown. The labels indicate whether the particle moves along ingoing ($I$) or outgoing ($O$) null paths with the first and second
subscripts respectively indicating the signs of $qQ$ and $E_\infty$. Physically the gray forbidden zones can be best understood as resulting from an electromagnetic redshifting of the wave 
vector when the particle moves against the field or a blueshifting when the particle moves with the field. In the gray zones the energy becomes negative (or equivalently the particles must 
move backwards in time).  
The dividing line between zones is always located at at $r= \frac{qQ}{E_\infty}$. }
\label{NullPaths}
\end{figure*}

Then possible particle paths are  shown in FIG.~\ref{NullPaths}. Intuitively they can be understood as the electromagnetic field redshifting or blueshifting $k^{\pm}$ depending on whether or 
not the particle is moving with or against the field\footnote{Thanks to Hari Kunduri for suggesting this interpretation.}. The energy vanishes at 
\be
r_o = \frac{qQ}{E_\infty^\pm} \, .  
\ee
That is, in order for the particle to have energy $E_\infty^\pm$ at infinity it must have zero energy at $r_o$. For particular choices of $q$, $Q$ and $E_\infty^\pm$ certain regions of
spacetime are forbidden to (future oriented) positive energy particles. 

Ori then argued that physically it makes more sense to view particles reaching $r_o$ as switching from ingoing to outgoing null paths rather than continuing in a straight line and so becoming
negative energy particles. Thus in FIG.~\ref{NullPaths} the ingoing particles in $I_{++}$ redshift to zero energy at $r_o$ and so bounce to become the outgoing particles of $O_{++}$. 
Similarly the outgoing particles of $O_{-+}$  bounce to become the ingoing particles of $I_{-+}$. 

This same interpretation may be applied to the particles making up the charged fluid in Vaidya RN. In that case the particles making up the shell of constant $v$ (or $u$) essentially move as if
they were particles of charge
\be
q = \frac{Q_v}{4 \pi r^2} 
\ee
moving in a background RN spacetime with mass $M(v)$ and charge $Q(v)$.

\section{Energy conditions for Type II stress-energy tensors}
\label{EC}

Stress-energy tensors are classified in \cite{hawking73} by their eigenvectors. For physical fields by far the most common are Type I tensors which have a timelike eigenvector $\xi^a$ 
whose eigenvalue is the (negative) energy density as measured by an observer with that four-velocity:
\be
T^a_{\phantom{a} b} \xi^b = - \mu \xi^a \, . 
\ee

However the focus of this paper is Type II tensors which have no timelike eigenvector but instead have a double null eigenvector. Then for some tetrad $(\ell, N, e^{(2)}, e^{(3)})$ where
$\ell$ and $N$ are null, future-oriented and cross-scaled so that $\ell \cdot N = -1$ and $e^{(2)}$ and $e^{(3)}$ are orthonormal (to each other) and orthogonal to $\ell$ and $N$, the stress-energy 
tensor  will necessarily take the form
\be
T^{ab} = \mu N^a N^b - \rho q^ {\perp ab} + P_{2} e_{(2)}^a  e_{(2)}^b + P_{3} e_{(3)}^a e_{(3 )}^b \label{TIIab} \, . 
\ee
Here $\mu \neq 0$ ($\mu = 0$ is Type I) and $q^\perp_{ab}$ is as defined in (\ref{qperp}). 
This particular arrangement of the constants has been chosen to be consistent with (\ref{NFstress}) though in  that case note that
$P_2 = P_3$. 

%

Then we can consider the restrictions placed on $\mu$, $\rho$, $P_{2}$ and $P_3$ by the energy conditions. The weak, dominant and strong conditions are each based on measurements of the 
stress-energy made by timelike observers. Thus  we consider an arbitrary future-oriented unit timelike vector field which can be defined by parameters $\alpha$, 
$\beta$ and $\gamma$:
\begin{align}
\xi^a = &\frac{\cosh \! \alpha}{\sqrt{2}} \left(e^\beta \ell^a + e^{-\beta} N^a \right)  \label{xi} \\
& + \sinh \! \alpha \left( (\cos \! \gamma) e_{(2)}^a + (\sin \! \gamma) e_{(3)}^a \right) \, . \nonumber
 \end{align}
 The null energy condition is based on an arbitrary null vector which we write similarly as
 \be
 k^a = \frac{e^\alpha}{\sqrt{2}} \left( e^\beta \ell^a + e^{-\beta} N^a \right) + e^\alpha \left(  (\cos \! \gamma) e_{(2)}^a + (\sin \! \gamma) e_{(3)}^a \right) \, . 
 \ee
 
 It is then straightforward to check the energy conditions. We present these in more detail than the complexity of the calculations
 might warrant as the results differ from those presented in \cite{hawking73}. While the correct energy conditions have been noted and applied 
 by others \cite{MMS1996,OriPC,Martin-Moruno:2013sfa,RefPC,Martin-Moruno:2017exc} it also true that the error in \cite{hawking73} does not seem to be universally known.
The incorrect conditions are used in, for example, 
 \cite{husain:1996,Wang:1998qx,Harko:2000ni,Ghosh2002,Debnath:2007vb,Ghosh:2008zza,Chatterjee:2015cyv}. 

 
 \subsection{Weak energy condition}
 The weak energy condition $T_{ab} \xi^a \xi^b \geq 0$ says that no timelike observer sees negative energy densities. From (\ref{TIIab}) and (\ref{xi}) this becomes
  \be
 \left(\frac{e^{2 \beta}}{2} \mu + \rho  \right) + (P_2 \cos^2 \!  \gamma + P_3 \sin^2 \! \gamma) \tanh^2 \!  \alpha \geq 0
 \ee 
for all $\alpha, \beta, \gamma$. By considering extreme cases we can find the bounds on $\mu$, $\rho$, $P_2$ and $P_3$. 
Then
\begin{align}
&\alpha = 0, \beta \rightarrow \infty \; \; \Longrightarrow \; \; \mu \geq 0 \\
&\alpha = 0, \beta \rightarrow - \infty \; \; \Longrightarrow  \; \; \rho \geq 0 
\end{align}
while 
\begin{align}
\alpha \rightarrow \infty, \beta \rightarrow - \infty 
\; \;  \Longrightarrow  \;  \;  \rho + P_i \geq 0 \nn 
\end{align}
for $i \in \{2,3\}$. Other limits are redundant. 

\subsection{Null energy condition}
The null energy condition replaces the timelike vector $\xi^a$ in the weak energy condition with $k^a$. That is
\be
e^{2 \beta} \mu + 2 \rho + 2 \left(P_2 \cos^2 \!  \gamma + P_3 \sin^2 \! \gamma \right) \geq 0 \, , 
\ee
where the $e^\alpha$ overall scaling of the null vector becomes irrelevant. Thus,
\begin{align}
& \beta \rightarrow \infty \; \; \Longrightarrow \; \; \mu \geq 0  \\
 & \beta \rightarrow -\infty \; \; \Longrightarrow \; \;  \rho + P_i \geq 0 
\end{align}
for $i \in \{2,3\}$. Other limits are redundant and in the usual way this is implied by the weak energy condition. 

\subsection{Dominant energy condition} 
The dominant energy condition says that $-T^a_{\phantom{a} b} \xi^a$ should be future directed and causal. That is, timelike observers should only see matter flowing forwards in time with speed
less than or equal to the speed of light. Future directed is ensured by
\be
 T_{a b} \xi^a \ell^b \geq 0 \;  \Longrightarrow  \; \mu e^\beta + \rho e^{-\beta} \geq 0 \;  \Longrightarrow  \;  \mu, \rho \geq 0  \label{future}
\ee
with the corresponding $N^a$ condition being redundant. 

Causality implies that $\| T_{ab} \xi^b \|^2  \leq 0$. This becomes
\be
\rho \left( \mu e^{2 \beta} + \rho \right) \cosh^2 \! \alpha - \left(P_2^2 \cos^2 \! \gamma + P_3 \sin^2 \! \gamma \right) \sinh^2 \! \alpha \geq 0  \, . 
\ee
The $\alpha = 0$ limit is redundant with (\ref{future}) however
\be
\alpha \rightarrow \infty , \beta \rightarrow - \infty \; \; \Longrightarrow \; \; 
  |P_i|  \leq | \rho | 
\ee
for $i \in \{2,3\}$. Other limits are redundant. 

\subsection{Strong energy condition}
The strong energy condition $R_{ab} \xi^a \xi^b \geq 0$ can be interpreted in a physical way but in essence is the geometric condition that must be assumed to 
prove results such as the singularity theorems. With our usual substitutions it becomes:
\begin{align}
0 \leq & \frac{1}{2} \left(e^{2 \beta} \mu + P_2 + P_3  \right)  \\
&  + \left(\rho - \frac{1}{2} (1 - 2 \cos^2 \! \gamma )P_2 - \frac{1}{2} (1- 2 \sin^2 \! \gamma)P_3  \right) \tanh^2 \! \alpha \, . \nn
\end{align}
Then 
\begin{align}
&  \alpha = 0, \beta \rightarrow \infty \; \; \Longrightarrow \; \; \mu \geq 0 \\
 & \alpha = 0, \beta \rightarrow -  \infty \; \; \Longrightarrow \; \; P_2 + P_3  \geq 0 
\end{align}
while 
\be
\alpha \rightarrow \infty, \beta \rightarrow - \infty  \; \;  \Longrightarrow \; \; \rho + P_i \geq 0 
\ee
for  $i \in (2,3)$. Other limits are redundant. 

\subsection{Summary of energy conditions}

To summarize, for a stress-energy tensor of form (\ref{TIIab}) the energy conditions are:
\begin{description}
\item[Weak]  $\mu \geq  0$ , $\rho \geq 0$  ,  $\rho + P_i \geq 0$  
\item[Null] $\mu \geq  0$,  $\rho + P_i \geq 0$  
\item[Dominant] $\mu \geq  0$ , $\rho \geq 0$ ,  $|P_i|  \leq | \rho |$
\item[Strong] $\mu \geq 0$, $P_2+P_3 \geq 0$, $\rho + P_i \geq 0$. 
\end{description}
If we restrict to Type II stress-energy tensors of this form then $\mu >  0$. 

As noted, individually these are not equivalent to the conditions given in \cite{hawking73}. However 
if $P_1=P_2$ and  we require all of them be satisfied simultaneously  then this is the same as requiring that all of the conditions in \cite{hawking73} be satisfied simultaneously. 
For anisotropic angular pressures ($P_1 \neq P_2$) the combined conditions are not quite equivalent as \cite{hawking73} also requires the pressures to be individually positive.

%
%

\section{Stress-energy is $C^1$ across $B$} 
\label{C1}

In this appendix we demonstrate that the stress-energy of polytropic fluids is not only continuous across $B$, the derivatives are also continuous. 
To see this we first derive the equations of motion governing the null fluid.
Either by expanding the divergence of (\ref{NFstress}) or (equivalently) by combining (\ref{mu})-(\ref{4}) it is straightforward to show that they are
\begin{align}
\Lie_{\! N} \! \left( \tilde{\varepsilon} \rho \right) + P \Lie_{\! N} \tilde{\varepsilon} & = 0 \; \; \; \; \mbox{and} \label{Neq}\\
\Lie_\ell \! \left( \tilde{\varepsilon} \rho \right) + P \Lie_\ell \tilde{\varepsilon} & = - \Lie_{\! N} (\tilde{\epsilon} \mu )  \label{Leq}
\end{align}
where $\tilde{\varepsilon} = r^2 \sin^2 \! \theta \mbox{d} \theta \wedge \mbox{d} \phi$ is the usual spherically symmetric area element. 
As always, these are conservation equations balancing evolving energy densities and work terms.

Now consider what these say about how the fields change across $B$. Writing the tangent and normal vectors as
\begin{align}
X &= \alpha \ell + \beta N  \; \; \; \;  \mbox{and}\\
X_{\! \perp} & = \alpha \ell - \beta N
\end{align}
the equations of motion (\ref{Neq}) and (\ref{Leq}) can be recast as
%
\begin{align}
& \Lie_{\! X} \! \left( \tilde{\varepsilon} \rho \right) + P \Lie_{\! X} \tilde{\varepsilon} =   \Lie_{\! X_{\! \perp}} \! \left( \tilde{\varepsilon} \rho \right) + P \Lie_{\! X_{\! \perp}} \tilde{\varepsilon} 
= \frac{\alpha}{2 \beta} \left(\Lie_{\! X_{\!\perp}} \! \! - \! \Lie_{\! X} \right) \left(\tilde{\varepsilon} \mu \right) \, . 
\end{align}
On $B$ with $f^+=f^-$ and $\mu = 0$ we saw in Section \ref{shell},  that intrinsic and extrinsic curvatures match and also $\rho^+ = \rho^-$. Then it immediately follows that
\begin{align}
 (\Delta P) \Lie_{\! X} \tilde{\varepsilon} & =   \tilde{\varepsilon} (\Delta  \Lie_{\! X_{\! \perp}} \! \rho)   + (\Delta P) \Lie_{\! X_{\! \perp}} \tilde{\varepsilon} 
  = \tilde{\varepsilon} \frac{\alpha}{2 \beta} \Delta \left(  \Lie_{\! X_{\!\perp}}  \mu \right)  
\end{align}
where $\Delta P = P^+ - P^-$ and similarly for the other quantities. Hence discontinuities in $P$ imply discontinuities in the normal derivatives of $\mu$ and $\rho$. 
However for polytropic models   $P = k \rho$  and so not only do $\mu$, $\rho$ and $P$ match across $B$ but so do their 
normal derivatives. 

By the matching conditions we already know that the tangential derivatives are continuous. Hence the derivatives of the stress-energy components are also continuous.

\bibliography{Null_Fluid_bib}

\end{document}